# Statistical mechanical study of thermodynamic properties of a family of fullerites from $C_{36}$ to $C_{96}$ in the equilibrium with their vapors

V.I.Zubov[*], I.V.Zubov and J.N. Teixeira Rabelo

*Instituto de Física, Universidade Federal de Goiás, C.P. 131, 74001-970, Goiânia, GO, Brasil[*]*
*and Russian Peoples' Friendship University, 117198, Moscow, Russia*

On the basis of the correlative method of the unsymmetrized self-consistent field that yields the account of the strong anharmonicity of the lattice vibrations, it has been calculated the temperature dependence of saturated vapor pressure of higher and smaller fullerites, from $C_{36}$ up to the $C_{96}$, and their thermodynamic properties along their sublimation curves. We have used the intermolecular potential of Girifalco with parameters recently calculated for these fullerenes. The calculations were accomplished up to the temperature of loss of stability (spinodal point) $T_s$. We compare our results with available experimental data, and with quantities calculated earlier for the magnitudes of the most widespread of the fullerites, the $C_{60}$. The behavior of some characteristics is considered in their dependence on the number of atoms in the molecule. The saturated vapor pressures up to the spinodal points of the two-phase systems crystal – gas is approximated by the formula $\log P_{sat} = A - (B/T) - CT$, where the last term is related to the anharmonicity of the lattice vibrations. The coefficient $A$ practically has no dependence on the number of atoms in the molecule (varying only by 2.2%), $B$ increases monotonically, while $C$ decreases from the $C_{36}$ to the $C_{96}$ by approximately twice. The isothermal bulk modulus $B_T$ and the shear modulus $C_{44}$ vanish at the spinodal points.

## 1. INTRODUCTION AND BRIEF SURVEY

The possibility of the existence of a third (along with the graphite and the diamond), molecular form of the carbon was theoretically investigated in the beginning of the seventies.[1,2] These molecules $C_{20+2n}$ (*n* is a positive integer number) have been experimentally discovered in the mid eighties.[3] Since the elaboration of effective technologies for the production, separation and purification of fullerenes in quantities enough for growing crystals of macroscopic dimensions, the fullerites,[4] it has been observed a permanent interest to their experimental and theoretical investigation. Hitherto, the most widespread $C_{60}$ and the next to it, $C_{70}$ are the most completely studied. It is known for instance that at low temperatures the molecules in their lattices are orientationally ordered, whereas at high temperatures they rotate rather freely in a fcc lattice (with a little mixture of the hcp phase in $C_{70}$). Similar behavior is to be expected also from other fullerites.

[*] To whom correspondence should be addressed. E-mail: zubov@fis.ufg.br

The theoretical study of the thermodynamics of the high-temperature modifications of the fullerites was initiated by Girifalco.[5] Considering that the form of the $C_{60}$ molecule is almost spherical and averaging the Lennard-Jones atom-atom potentials of a pair of molecules over all their orientations, he has deduced for the orientationally disordered phases (solid, gaseous and hypothetical liquid) the intermolecular potential

$$\Phi_G(r) = -\alpha \left( \frac{1}{s(s-1)^3} + \frac{1}{s(s+1)^3} - \frac{2}{s^4} \right) + \beta \left( \frac{1}{s(s-1)^9} + \frac{1}{s(s+1)^9} - \frac{2}{s^{10}} \right). \tag{1}$$

Here $s = r/2a$, $r$ is the distance between the centers of the molecules; $a$ is the radius of their hard core,

$$\alpha = n^2 A / 12(2a)^6, \quad \beta = n^2 B / 90(2a)^{12}, \tag{2}$$

$A$ and $B$ are the coefficients to the attractive and repulsive terms of the atom-atom Lennard-Jones potential, while $n = 60$ is the number of atoms in the molecule. The parameters $A$ and $B$ were fitted in with experimental data for the lattice constant and heat of sublimation.

The $C_{70}$ molecule has a form similar to an oblong ellipsoid. The forms of molecules of the other fullerenes are even more complicated. Noticing, that this molecule can be separated in five groups of 10 or 20 atoms, each one lying in a spherical shell of a certain radius, and generalizing the procedure of Girifalco, Verheijen et al.[6] obtained an intermolecular potential for this fullerene. It has the form of a sum of 25 terms of the type (1), describing the interactions between different pairs of such shells of neighboring molecules.

Kniaz′, Girifalco and Fischer, [7] and independently, Abramo and Caccamo[8] used the Girifalco potential (1) for the orientationally disordered phases of $C_{70}$. This corresponds to approximate the form of the molecule to a sphere, whose radius is determined by fitting in the calculated with this potential lattice constant with its experimental value. The idea of a spherical approximation for some of the higher fullerenes with radius related to the number of atoms in the molecule was used in Refs. 9, 10 but regardless to the intermolecular potentials.

The intermolecular potentials of Girifalco and Verheijen were employed on the basis of the correlative method of the unsymmetrized self-consistent field (CUSF, see e.g. Refs. 11, 12) for the investigation of the whole sets of equilibrium thermodynamic properties of the high-temperature phases of the fullerites $C_{60}$ [11 - 13] and $C_{70}$.[14, 15] Broad regions of their diagram of states were considered, including the sublimation curves.[12, 14] The intra-molecular vibrations[16 -18] was taken into account. They give the main contribution to the specific heats. The agreement with experiment is good.

In the last years, a growing interest has been observed to the higher fullerites $C_{76}$ and $C_{84}$,[19 - 25] and to $C_{96}$,[26] and also to the smaller ones, specially to $C_{36}$.[27, 28] A method was proposed[29] for the calculation of the coefficients of the potential of Girifalco (1) basing on the spherical approximation of the form of the molecules, starting from their magnitudes for the $C_{60}$,[5] i.e. without additional fitting parameters. There have been calculated the coefficients for a series of smaller and higher fullerenes, from the $C_{28}$ to the $C_{96}$. The curves of the potentials for these fullerenes are shown in Fig.1, and the refined values of their basic characteristics are presented in Table I. It is interesting that the coefficients $\alpha$ and $\beta$ decrease with increasing number of atoms in the molecule, although the minimum point of the potential and the depth of its well of course increase. One can see from Fig. 1 that the minimum points of the Girifalco potential for various fullerenes lie on a nearly straight line. This potential has been used for the calculation of the saturated vapor pressures and the thermodynamic properties of the two higher fullerites $C_{76}$ and $C_{84}$.[30] A good agreement with available experimental data has been obtained.

Recently, computer simulation results have been published for the properties of $C_{76}$ and $C_{84}$ fullerites using the Girifalco potential.[31, 32] Fernandes et al.[31] have utilized in particular our values of the parameters for this potential[29, 30] whereas Micali et al.[32] adopted the effective radii of the molecule by fitting experimental data for the lattice parameters. Both approaches give close results.

The present work is devoted to theoretical investigation and comparison analysis of thermodynamic properties of a family of fullerites from $C_{36}$ to $C_{96}$ along their sublimation curves.

## 2. BRIEF SKETCH OF THE METHOD

The zeroth approximation of the CUSF includes the strong anharmonicity of the lattice vibrations up to the fourth order, and the perturbation theory takes into account the fifth and sixth anharmonic terms. Thus taking into consideration the rotational and intra-molecular degrees of freedom, the free Helmholtz energy and the equation of state of the crystal at a temperature $\Theta = kT$ under a pressure $P$ are of the form

$$
\begin{aligned}
F = N \Bigg\{ & \frac{K_0}{2} - \frac{5\Theta}{24} \left( \frac{\beta}{X} \right)^2 - \frac{\Theta}{4} \left( X + \frac{5\beta}{6X} \right)^2 \\
& - \Theta \ln \left[ \left( \frac{3m^2\Theta^3}{\hbar^4 K_4} \right)^{3/4} D_{-1.5} \left( X + \frac{5\beta}{6X} \right) \right] \Bigg\} - \frac{N\Theta}{2} \ln \left( \frac{8 I_1^2 I_2 \Theta^3}{\hbar^6} \right) \\
& - N\Theta \sum_j g_j \ln \left( 2 \sinh \frac{\hbar \omega_j}{\Theta} \right) + F^2 + F^H ;
\end{aligned}
\tag{3}
$$

$$
P = -\frac{a}{3v} \left[ \frac{1}{2} \frac{dK_0}{da} + \frac{\beta}{2K_2} \frac{\Theta}{da} \frac{dK_2}{da} + \frac{(3-\beta)\Theta}{4K_4} \frac{dK_4}{da} \right] + P^2 + P^H ;
\tag{4}
$$

$$
E = \frac{N}{2} \left[ K_0 + \frac{(15+\beta)\Theta}{2} + \frac{\hbar}{2} \sum_j g_j \omega_j \coth \frac{\hbar \omega_j}{2\Theta} \right] \\
+ E^2 + E^H
\tag{5}
$$

Here $N$ is the quantity of molecules (the Avogadro's number), $m$ the mass of the molecule, $I_1$ and $I_2$ are the moments of inertia (for $C_{60}$ $I_1 = I_2$), $\omega_j$ and $g_j$ are the frequencies and degeneracies of the intra-molecular vibrations, $a$ is the distance between the nearest neighbors, $v(a) = V/N$ the volume of the unitary cell, and $\beta\left( K_2(3/\Theta K_4)^{1/2} \right)$ is an implicit function determined by the transcendental equation

$$
\beta = 3X \frac{D_{-2.5}(X + 5\beta/6X)}{D_{-1.5}(X + 5\beta/6X)} ,
\tag{6}
$$

where $D_\nu$ are the parabolic cylinder functions;

$$
K_{2l} = \frac{1}{2l+1} \sum_{k \geq 1} Z_k \nabla^{2l} \Phi(R_k), \quad l = 0, 1, 2 ,
\tag{7}
$$

$F^2$, $F^H$, $P^2$ and $P^H$ are the corrections of the perturbation theory that contain in particular the anharmonicity of higher orders. It is seen from eq 4, for instance, that the rotational and intra-molecular degrees of freedom (that the prevalent to the specific heats of the fullerites) have no influence on the equation of state and on the properties related to it, the thermal expansion and the isothermal elastic moduli. But they give a significant contribution to the equation of energy (eq 5) and the main contribution to the specific heats of the fullerites (up to 80 – 90 %). Because of this, the molar specific heats of the fullerites at high temperatures are much greater than those of the majority of other substances, and the difference between isobaric and isochoric specific heats as well as between isothermal and adiabatic elastic moduli are very small.

In the case of equilibrium of the crystal with its vapor one has to add to the eq 4 the condition of phase equilibrium (equality of the chemical potentials) and the equation of state of the gaseous phase. For the latter one can use the virial expansion. Taking into account the second virial terms this reduces to

$$P = P_{id}\left(1 - BP_{id}/\Theta\right);$$

$$P_{id} = \Theta\left(\frac{K_4}{12\pi^2\Theta}\right)^{3/4}\exp\left[\frac{K_0}{2\Theta} - \frac{5}{24}\left(\frac{\beta}{X}\right)^2 - \frac{1}{4}\left(X + \frac{5\beta}{6X}\right)^2 + \frac{f^2 + f^H}{\Theta}\right]\bigg/ D_{-1.5}\left(X + \frac{5\beta}{6X}\right), \quad (8)$$

where $f^i = F^i/N$, $i = 2, H$. The system of equations 4 and 8, with the account of 6, determine the temperature dependence of the saturated vapor pressure $P_{sat}$ $(T)$ and of the distance between the nearest neighbors in the crystal $a$ $(T)$ along the curve of phase equilibrium.

## 3. RESULTS AND DISCUSSION

**3.1. Sublimation curves**. We have solved these equations for several fullerites from $C_{36}$ to $C_{96}$. The results for the mean intermolecular distances in $C_{36}$, $C_{50}$, $C_{60}$, $C_{76}$ and $C_{96}$ together with the available experimental data for $C_{60}$[33, 34] and $C_{76}$[19] are depicted in Fig. 2. Remember [12] that the upper branches $a_2(T)$ of these curves correspond to the absolute unstable thermodynamic states, since on them the isothermal modulus of the crystal $B_T$ is negative. At the temperature $T_S$, where both branches coalesce $B_T$ vanishes. From here it is seen the good agreement for the lower branches $a_1(T)$ with available experimental data. It is interesting that the points $a(T_S)$ lie practically on a straight line.

To avoid encumbrance, the dependence of the logarithm of the saturated vapor pressures on the inverse of temperature is demonstrated in Figure 3 only for $C_{36}$, $C_{60}$ and $C_{96}$. It is close to linear. For the lower branch this is in good agreement with available results of measurements for $C_{60}$[35 - 38] (and also for $C_{70}$, $C_{76}$ and $C_{84}$). More accurately, along all the range of temperatures it is described by the equation

$$\log P_{sat} = A - \frac{B}{T} - CT, \quad (9)$$

in which the last term is due to the anharmonicity of lattice vibrations. Although the coefficient $C$ is relatively small, yet at high temperatures it gives a noticeable contribution to $P_{sat}$. We have processed by this formula the results of our calculations up to the temperatures of the possible triple points crystal – liquid – vapor. The existent estimates for the $C_{60}$ lie in the limits from 1450[12] to 1800 K.[39] We have taken the mean value 1620 K,[40] and for the other fullerites we have used the rule of the correspondent states. The behavior of all the three coefficients is shown in Figure 4. The constant term in eq 9 depends almost not at all on the number of atoms in the molecule, varying only by 2.2%. Thus in the spinodal points, the saturated vapor pressures of various fullerites differ very little from each other. The coefficient $B$ grows monotonically, while $C$ decreases from the $C_{36}$ to the $C_{96}$, by about twice. Such a behavior agrees with experimental estimations of $A$ and $B$ available for $C_{60}$, $C_{70}$, $C_{76}$ and $C_{84}$, see Table 2. There is a detour from this trend of the value of $B$ for the $C_{70}$, calculated using the Verheijen intermolecular potential.

All the data of measurements executed for $C_{60}$, $C_{70}$, $C_{76}$ and $C_{84}$, due to experimental difficulties in limited temperature intervals have been processed in the linear approximation, i.e. without the third term in eq 9. Markov, *et al.*[25] based on the analysis of the results of various authors for $C_{60}$ and $C_{70}$ have given values recommended of the coefficients $A$ and $B$ for these fullerites. In Table 2 we compare our results with them and with experimental data for $C_{76}$ [20-22] and for $C_{84}$.[22–24]

In Figure 5 we show the temperature dependence of the enthalpies of sublimation of fullerites. We compare them with the experimental data for $C_{60}$ and $C_{70}$[25], $C_{76}$[20, 21], and $C_{84}$.[23, 24]

One can see the agreement within the limits of experimental error, with the exception of the results for $C_{76}$[20] whose difference from our calculations slightly extrapolated these limits. Our results show an excellent agreement with computer simulations by Fernandes et al.[32] presented at 700 K (in kJ/mole): $170\pm12$ for $C_{60}$, $191\pm13$ for $C_{70}$, $198\pm14$ for $C_{76}$, and $212\pm15$ for $C_{84}$.

**3.2. Thermodynamic properties.** Along the lower branches we have calculated the thermodynamic properties of the fullerites that are due to the lattice vibrations.

Figures 6 and 7 show the elastic properties, which are also the coefficients of thermodynamic stability of the crystals. They all decrease monotonically as the temperature increases. Besides, in the heavy fullerites they are greater than in the light ones. At $T = T_S$ the bulk modulus $B_T$ vanishes together with the shear modulus $C_{44}$, both as the ½ power of $T_S - T$. Other stability coefficients remain finite and positive. Therefore, $T_S$ is the spinodal point (the point of the loss of the thermodynamic stability), in the present case for the two-phase system crystal - vapor.

In Figure 8, the thermal expansion coefficient of fullerites is shown. It increases with the temperature. For the heavy fullerites it is smaller than for the light ones. Near the spinodal point it grows sharply tending to infinite when $T \to T_S$. That is why, we take here a smaller temperature interval.

The spectra of intra-molecular vibrations $\omega_j$ and $g_j$ that give the prevalent contribution to the specific heats of the fullerites are known for $C_{60}$ and $C_{70}$ fullerenes.[16 - 18] That's why, we have calculated the thermal properties only for this two fullerites. Their isobaric specific heats are shown in Fig. 8. Their molar values are considerably greater than those of the majority of other substances. Besides for $C_{70}$ it is higher than for $C_{60}$ because of the greater number of its intra-molecular degrees of freedom. At the same time, the gram atom specific heats of both fullerites are very close to each other (except for the small vicinity of the points $T_S$, where the contribution of the lattice vibrations grows sharply). This is an indication of the proximity of the integral characteristics of their intra-molecular vibrations. The agreement with experimental results available for $C_{60}$ fullerite[25, 41 - 44] is quite satisfactory.

So, the good agreement with experimental data that we have obtained for thermodynamic properties of fullerites $C_{60}$[11-13] and $C_{70}$,[14,15] and also for sublimation characteristics of $C_{76}$ and $C_{84}$[30, 31] allows us to expect an agreement of results of our present calculations for other fullerites with those of measurements that will appear in the future.

**Acknoeledgment.** This work was supported financially in part by Conselho Nacional do Desenvolvimento Científico e Tecnológico – CNPq (Brazil). V.I.Zubov is grateful for this support.

Table 1. Characteristics of the Girifalco potential (1) for different fullerenes

| | $C_{60}$[a] | $C_{28}$ | $C_{36}$ | $C_{50}$ | $C_{70}$ | $C_{76}$ | $C_{84}$ | $C_{96}$ |
|---|---|---|---|---|---|---|---|---|
| $2a$ ($10^{-8}$ cm) | 7.100 | 4.850 | 5.500 | 6.481 | 7.669 | 7.991 | 8.401 | 8.981 |
| $\alpha$ ($10^{-14}$ erg) | 7.494 | 16.059 | 12.49 | 8.980 | 6.424 | 5.916 | 5.353 | 4.684 |
| $\beta$ ($10^{-17}$ erg) | 13.595 | 286.64 | 104.90 | 28.19 | 8.338 | 5.281 | 3.539 | 2.074 |
| $\sigma$ ($10^{-8}$ cm) | 9.599 | 7.358 | 8.000 | 8.976 | 10.161 | 10.480 | 10.880 | 11.468 |
| $r_0$ ($10^{-8}$ cm) | 10.056 | 7.808 | 8.460 | 9.443 | 10.622 | 10.946 | 11.358 | 11.927 |
| $-\varepsilon/k_B$ (K) | 3218 | 1775 | 2182 | 2813 | 3594 | 3807 | 4080 | 4467 |

[a] Girifalco.[5]

Table 2. Coefficients of Eq 9

| | $C_{60}$ | $C_{70}$ | $C_{76}$ | $C_{84}$ |
|---|---|---|---|---|
| $A^a$ | 9.3805 | 9.3975 | 9.4078 | 9.4115 |
| $A^b$ | 10.85±0.76 | 11.23±1.49 | - | - |
| $A^c$ | - | - | 11.23±0.20 | 10.92±0.30 |
| $A^d$ | - | - | 10.65±0.47 | 10.92±0.28 |
| $A^e$ | - | - | 10.66±0.92 | 11.28±0.93 |
| $B^a$ | 9019.6 | 9983.0 | 10533.8 | 11227.8 |
| $B^b$ | 8738±472 | 9768±774 | - | - |
| $B^c$ | - | - | 10150±150 | 10950±300 |
| $B^d$ | - | - | 9904±356 | 10570±234 |
| $B^e$ | - | - | 11365±709 | 11743±717 |

[a] Our results.

[b] Recommended experimental values obtained by Markov et al.[25] for $C_{60}$ in the range of 660 to

  990 K and for $C_{70}$ of 650 to 904 K.

[c] Experimental values obtained by Brunetti et al.[20] for $C_{76}$ in the range of 834 to 1069 K and by

  Piacenti et al.[23] for $C_{84}$ in the range of 920 to 1190 K.

[d] Experimental values obtained by Boltanina et al.[21] for $C_{76}$ in the range of 637 to 911 K and by

  Boltanina et al.[21] for $C_{84}$ in the range of 658 to 980 K.

[e] Experimental values obtained by Boltanina et al.[24] for $C_{76}$ and $C_{84}$ in the range of 681 to

  901 K.

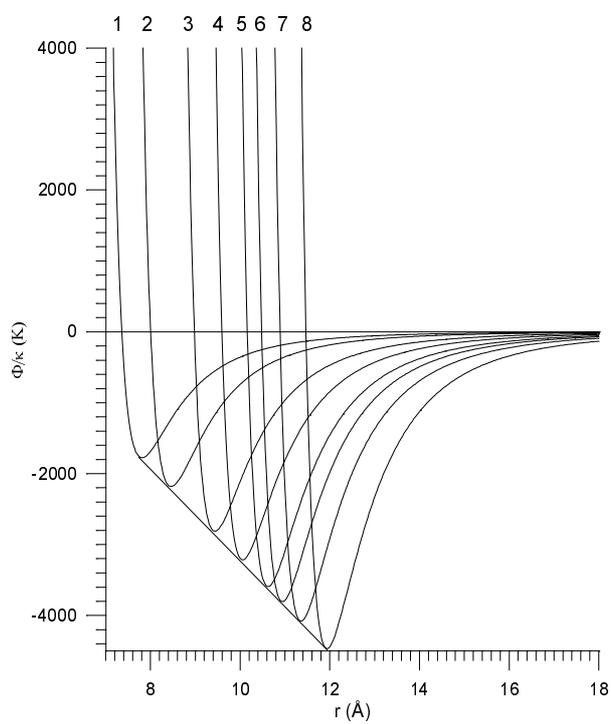

Figure 1. The Girifalco potential for eight fullerenes: $1 - C_{28}$, $2 - C_{36}$, $3 - C_{50}$, $4 - C_{60}$,[5] $5 - C_{70}$, $6 - C_{76}$, $7 - C_{84}$, $8 - C_{96}$.

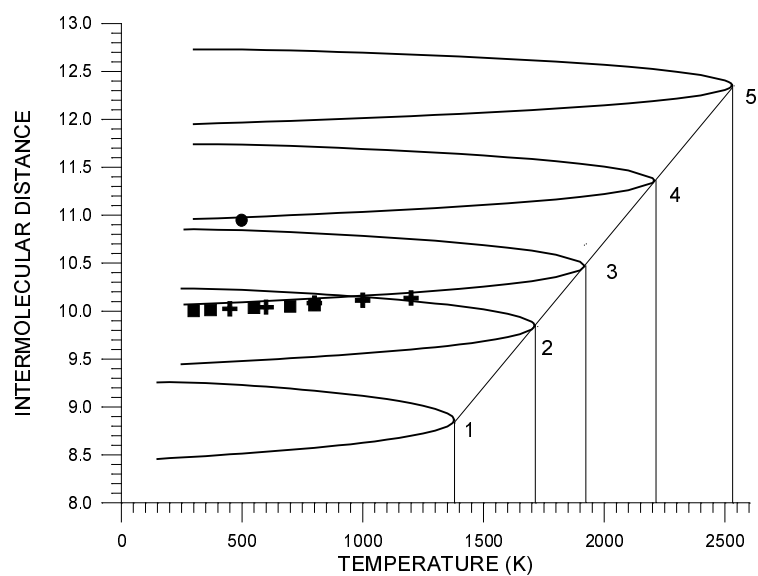

Figure 2. The intermolecular distances in the fullerites along their sublimation curves: $1 - C_{36}$, $2 - C_{50}$, $3 - C_{60}$, $4 - C_{76}$, $5 - C_{96}$. The vertical lines show the temperatures $T_S$. Experimental data for $C_{60}$ are taken from works of Mathews *et al.*[33] (■) and of Fischer and Heiney[34] (+), and for $C_{76}$ from work of Kawada *et al.*[19].

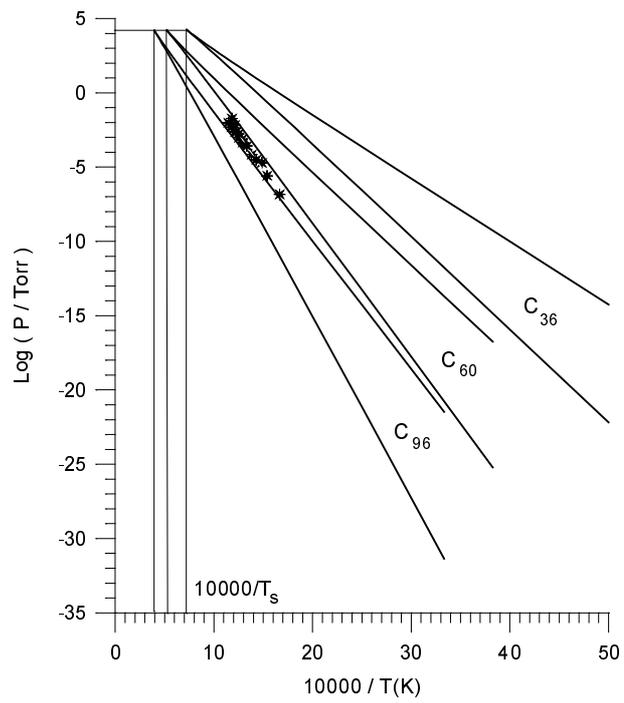

Figure 3. Saturated vapors pressures. Experimental data taken from Refs. 35 - 38.

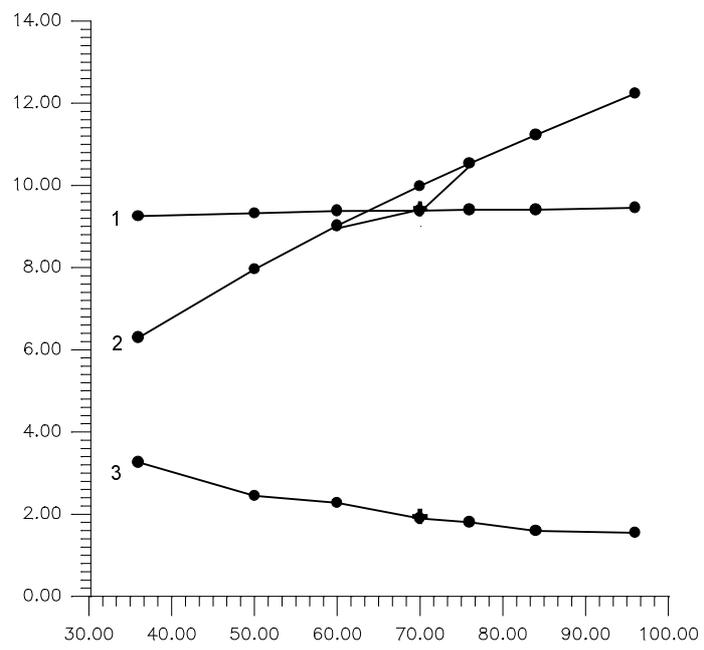

Figure 4. Coefficients of Eq 9: $1 - A$, $2 - 10^{-3}B$, $3 - 10^{4}C$.

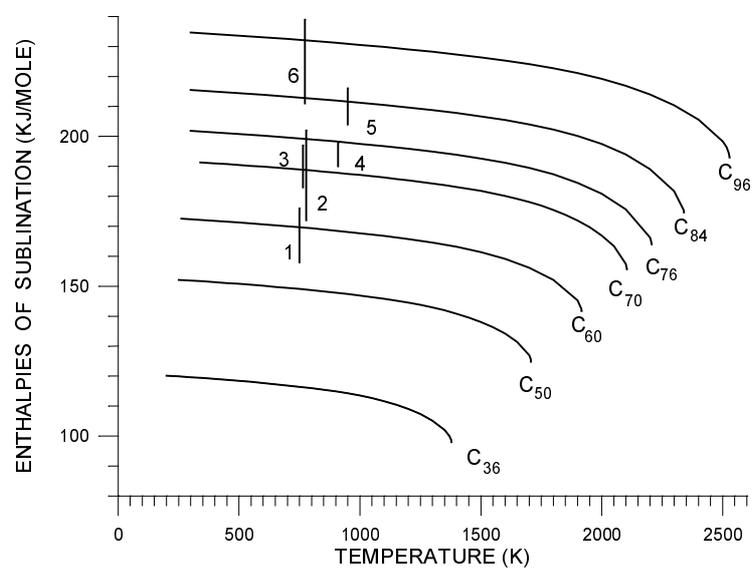

Figure 5. Enthalpies of sublimation. Experimental data for $C_{60}$ (1) and $C_{70}$ (2) are taken from work by Markov, et al.[25] (the values recommended), for $C_{76}$ from works by Brunetti et al.[20] (3) and by Boltanina et al.[24] (4), and for $C_{84}$ from works by Piacente et al.[23] and Boltanina et al.[24].

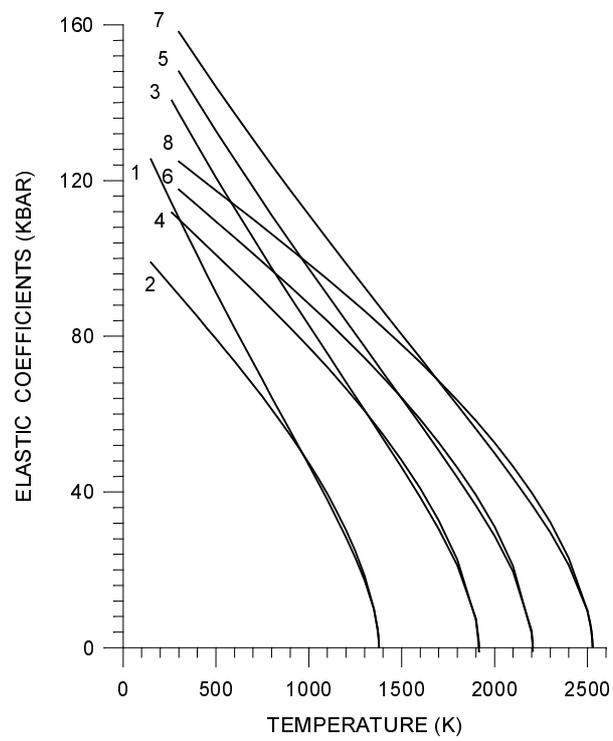

Figure 6. Isothermal bulk moduli $B_T$ (odd numbers) and shear coefficients $C_{44}$ (even numbers) of the fullerites: $1, 2 - C_{36}$, $3, 4 - C_{60}$, $5, 6 - C_{76}$, $7, 8 - C_{96}$.

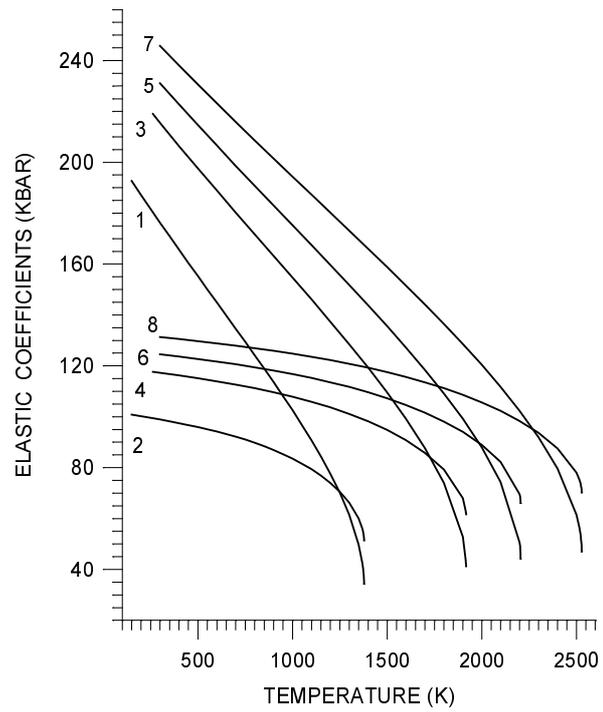

Figure 7. Isothermal coefficients of uniaxial tension $C^{T}_{11}$ (odd numbers) and shear coefficients $C^{T}_{11}$ - $C^{T}_{12}$. The numbers for various fullerites are the same as in Figure 5.

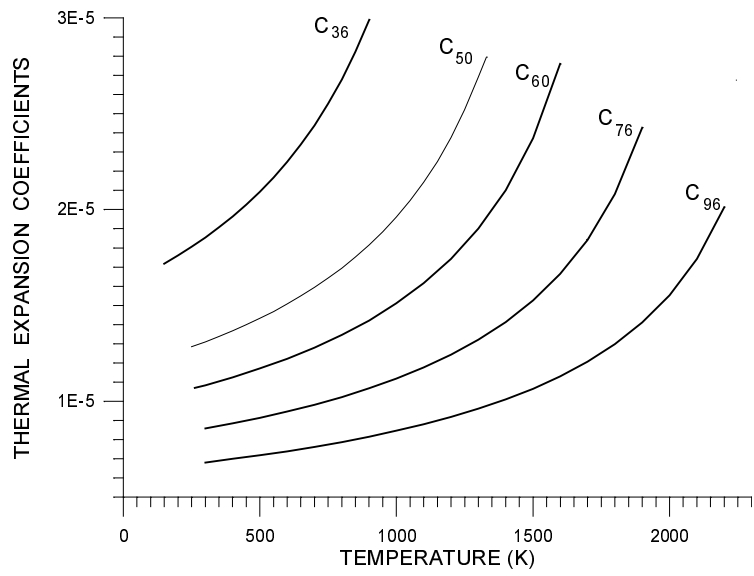

Figure 8. Thermal expansion coefficients of the fullerites.

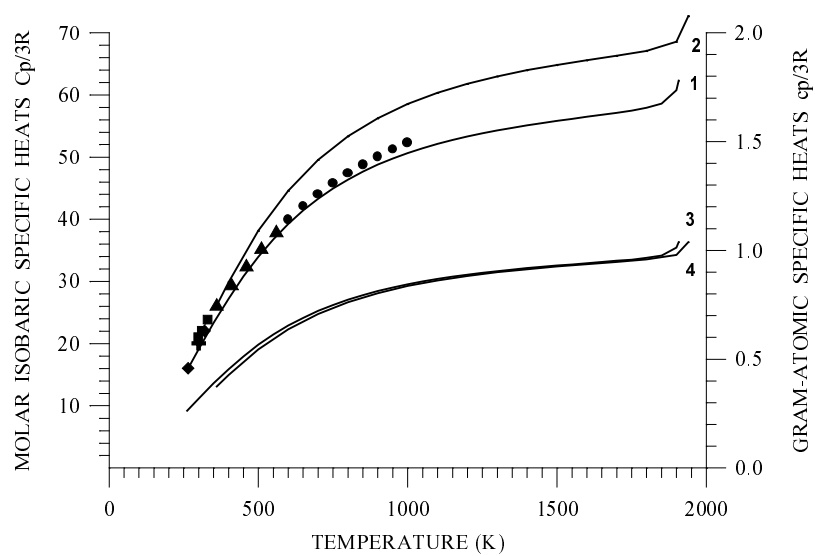

Figure 9. Isobaric heat capacities of C$_{60}$ and C$_{70}$. Experimental data are taken from works by Y.Jin *et al.*[41] (■), Matsuo *et al.*[42] (✚), Fischer *et al.*[43] (♦) , Lebedev *et al.*[44]▲) and Markov *et al.*[25] (●)